\documentstyle[preprint,eqsecnum,aps,psfig]{revtex}

\begin{document}
\draft
\preprint{HEP/123-qed}
\title{Photoionization cross sections of O II, O III, O IV, and O V:
benchmarking R-matrix theory and experiments}
\author{Sultana N. Nahar}
\address{
Department of Astronomy, The Ohio State University, Columbus, OH 43210\\
}
\date{\today}
\maketitle
\begin{abstract}
For crucial tests between theory and experiment, ab initio close 
coupling calculations are carried out for photoionization of several 
oxgyen ions: O II, O III, O IV, O V. The relativistic fine structure and 
resonance effects are studied using the R-matrix and its relativistic 
variant the Breit Pauli R-matrix (BPRM) approximations. Extremely detailed
comparison is made with high resolution experimental measurements carried
out in three different set-ups: Advanced Light Source at Berkeley, 
synchrotron radiation experiment with a new modulator at University of 
Aarh\"{u}s, and synchrotron radiation experiment at University of 
Paris-Sud. The comparisons illustrate physical effects in
photoionization such as (i) fine structure, (ii) resolution, and (iii)
metastable components. Photoionization cross sections $\sigma_{PI}$
of the ground and a few low lying excited states of these ions obtained 
in the experimental spectrum include combined features of these states. 
The general features of the measured $\sigma_{PI}$ were predicted in the 
earlier calculations in LS coupling. However, due to higher resolution 
achievement in the recent experiments, the theoretically calculated 
resonances need to be resolved with much finer energy mesh for precise 
comparison. In addition, prominent resonant features are observed in 
the measured spectra that can form from transitions allowed with
relativistic fine structure, but not in LS coupling. Present results
include calculations from both LS coupling and the relativistic BPRM
method to decipher the features in photoionization spectra. 
$\sigma_{PI}$ are obtained for ground and metastable (i) 
$2s^22p^3(^4S^o,^2D^o, ^2P^o)$ states of O II, (ii) 
$2s^22p^2(^3P,^1D,^1S)$ and $2s2p^3(^5S^o)$ states of O III, (iii) 
$2s^22p(^2P^o_J)$ and $2s2p^2(^4P_J)$ levels of O IV, and (iv) 
$2s^2(^1S)$ and $2s2p(^3P^o,^1P^o)$ states of O V. Most of the 
computed theoretical resonance structures are seen experimentally,
although there are some discrepancies. One of the main conclusions of 
the present study is that resonances in
photoionization cross sections of the ground and metastable states can
be a diagnostic of experimental beam composition, with potential 
applications to astrophysical and laboratory plasma environments. 
\end{abstract}
\pacs{PACS number(s):  32.80.Fb}

\narrowtext

\section{INTRODUCTION\protect\\}
\label{sec:level1}

Photoionization cross sections of many atoms and ions were studied in 
detail with resonance structures for ground and many excited states
under the Opacity Project (OP, \cite{op}). The close coupling (CC) 
R-matrix method, as employed under the OP, accurately considers the 
numerous autoionizing resonances in cross sections along interacting 
(overlapping) Rydberg series. The study under the OP resulted in new 
features in the cross sections, such as prominent PEC (photo-excitation
of core) resonances. The radiative work under the OP has been 
extened to collisional and radiative processes under the Iron Project 
(IP, \cite{ip}) with inclusion of relativistic effects using the BPRM
method. However, while the methods used are state-of-the-art, possible 
deficiencies are: (i) the resonance structures in the OP/IP cross 
sections are not always fully resolved, and (ii) the important
near-threshold resonances may not be accurate owing to inadequate
eigenfunction expansions.
A prime need therefore is to benchmark the vast amount of theoretical
OP/IP data that is now being utilized for global applications such as
calculation of stellar opacities, photoionization rates in astrophysical
models.

On the experimental side, accurate photoionization cross sections are 
being measured at several places, such as the merged ion-photon beam 
facility at the University of Aarh\"{u}s (e.g. for C II \cite{kjl}), the 
Electron Cyclotron Resonant Ion Source (ECR) at the University of 
Paris-Sud (e.g. for Xe V - Xe VIII \cite{fw}), and the Advanced Light 
Source (ALS) at Berekely (e.g. for O II \cite{als}). Each of these three 
sophisticated experimental set-ups has different advantages in 
measurements, and the results often complement each other.
Cross sections for O II have been measured at ALS for near threshold
features in $\sigma_{PI}$ of metastable states \cite{als}, including a 
wider photon energy range \cite{o2a}, 
and at the University of Aarh\"{u}s \cite{o2b}. Detailed 
photoionization cross sections have recently been measured for the 
ground and low excited states of multiply charged oxygen ions for the
first time for O III, O IV, and O V, at the synchrotron radiation source 
at University of Paris-Sud \cite{oions}. Study of oxygen ions is of great 
importance because of their abundances in terrestrial and
planetary atmospheres and in astrophysical objects. 

Comparison of all these results is extremely crucial to evaluate the 
accuracy of both the theoretical and experimental results, and to study 
the resonance phenomena. Theoretical work can also provide spectroscopic 
identification of resonances, and a quantitative measure of the mixture 
of states in the experimental beam. The resonant structures and the 
background shapes of the recently measured photoionization cross sections 
of oxygen ions are found to be in close agreement overall with the some 
ealier theoretical cross sections \cite{snn1}. 
However, certain features in the measured cross sections are not seen 
prominently or are missing in the calculated cross sections. 

One of the aims of this study is to study relativistic effects and fine
structure. We distinguish these in terms of relativistic mixing and fine
structure splitting with reference to pure LS coupling results. 
While relativistic effects per se might not be of great
importance for these light ions, fine structure splitting 
manifests itself in many experimentally observed features. In addition,
relativistic mixing via E1 transitions (e.g. singlet-triplet, 
doublet-quartet) results in additional resonances not allowed in LS
coupling. The present study investigates the LS coupling and fine 
structure effects together with resonance phenonmena in detail. Thus to 
begin with it is necessary to do LS coupling calculations at high 
resolution. Comparison with experiment then reveals whether fine 
structure is also to be considered in order to analyze experimental data. 
In such cases the relativistic BPRM method is employed to compute cross 
sections.

Detailed comparison of experimental and theoretical resonance positions 
and shapes also 
highlights the importance of correlation effects in theoretical
calculations. Resonances may be shifted with respect to their respective
thresholds of convergence, indicating missing correlation configurations
in the wavefunction expansion. This again is crucial since
the inherent accuracy of the close coupling calculations depends on the
completeness of the eigenfunction expansions, in particular short-range
correlation at low energies.

Another motivation for these studies is that the R-matrix method, as
developed for the OP/IP work, has been extended to a theoretically 
self-consistent treatment of photoionization and unified electron-ion 
recombination including radiative and dielectronic recombination in an 
ab initio manner\cite{np,znp,petal}. The total recombination cross 
sections have been benchmarked for accuracy by comparing with several 
experimental cross sections \cite{znp,petal}. But while recombination 
cross sections require total contributions of photorecombination cross 
sections from all bound 
levels, obtained from the correnpoiding $\sigma_{PI}$ using detailed
balance, comparison of photoionization cross sections with experiment 
involves individual features of $\sigma_{PI}$ of single states. For
instance,
photoionization cross sections including relativistic effects in the 
close coupling approximation, as developed under the IP, was first 
benchmarked by comparing with the measured $\sigma_{PI}$ of C II 
\cite{snn2}.

We outline the theory and calculations in Sections II and III
respectively, followed by a discussion of results for the four ions
O~II~-~O~V in Section IV, and the conclusions in Section V.

\section{THEORY \protect\\}
\label{sec:level2}

Present calculations are carried out in an {\it ab initio} manner using
the close coupling R-matrix method (e.g. \cite{mjs,op2}) as in the
Opacity Project \cite{op} and the Iron Project \cite{ip}. The 
package of codes from the OP \cite{op2} have been extended to include 
relativistic effects in the Breit-Pauli approximation in the IP work
\cite{ip,st,ip2} under the name Breit-Pauli R-matrix (BPRM) method. A
brief discussion of the method is given below.

In the close coupling approximation the total wavefunction of 
the (N+1)-electron ion is represented by the wavefunctions of the 
N-electron core, multiplied by the wavefunction of the outer electron:

\begin{equation}
\Psi_E(e+ion) = A \sum_i \chi_i(ion)\theta_i + \sum_{j} c_j \Phi_j(e+ion),
\end{equation}

\noindent
where $\chi_{i}$ is the target (or the core) wavefunction in a specific 
state $S_iL_i\pi_i$ or level $J_i\pi_i$ and $\theta_{i}$ is the 
wavefunction for the ($N$+1)-th electron in a channel labeled 
$S_iL_i(J_i)\pi_ik_{i}^{2}\l_i(\ J\pi)$; $k_{i}^{2}$ is its incident 
kinetic electron energy. $\Phi_j$'s are the correlation functions of the 
($N$+1)-electron system that account for short range correlation and the 
orthogonality between the continuum and the bound orbitals. The complex 
resonant structures in photoionization are included through channel 
couplings.

Relativistic effects are included through Breit-Pauli approximation in
intermediate coupling. The Breit-Pauli Hamiltonian, as adopted in the IP
work \cite{ip}, has the following terms, 
\begin{equation}
H_{N+1}^{\rm BP}=H_{N+1}+H_{N+1}^{\rm mass} + H_{N+1}^{\rm Dar}
+ H_{N+1}^{\rm so},
\end{equation}
where $H_{N+1}$ is the nonrelativistic Hamiltonian,
\begin{equation}
H_{N+1} = \sum_{i=1}\sp{N+1}\left\{-\nabla_i\sp 2 - \frac{2Z}{r_i}
        + \sum_{j>i}\sp{N+1} \frac{2}{r_{ij}}\right\},
\end{equation}
and the additional terms represent the one-body mass correction,
the Darwin, and the spin-orbit interaction terms, respectively:
\begin{equation}
H_{N+1}^{\rm mass} = -{\alpha^2\over 4}\sum_i{p_i^4}, ~
H_{N+1}^{\rm Dar} = {Z\alpha^2 \over 4}\sum_i{\nabla^2({1\over r_i})}, ~
H_{N+1}^{\rm so}= Z\alpha^2 \sum_i{1\over r_i^3}{\bf l_i.s_i}.
\end{equation}

The set of ${SL\pi}$ are recoupled to obtain (e + ion) states  with
total $J\pi$, following the diagonalization of the (N+1)-electron
Hamiltonian to solve
\begin{equation}
H^{BP}_{N+1}\mit\Psi_E = E\mit\Psi_E.
\end{equation}
Substitution of the wavefunction expansion introduces a set of coupled
equations that are solved using the R-matrix approach. The continuun
wavefunction, $\Psi_F$, describe the scattering process with the free
electron interacting with the target at positive energies (E $>$ 0),
while the solutions correspond to pure bound states $\Psi_B$ at 
{\it negative} total energies (E $<$ 0).

The photoionization cross section ($\sigma_{PI}$) is proportional to
the generalized line strength ($S$),
\begin{equation}
\sigma_{PI} = {4\pi \over 3c}{1\over g_i}\omega S.
\end{equation}
where $S$ is the generalized line strength:
\begin{equation}
S_{\rm L}= |<\Psi_j||{\bf D}_L||\Psi_i>|^2,
\end{equation}
{\bf D} is the dipole operator; ${\bf D}_L = \sum_i{r_i}$ in "length" form, 
and ${\bf D}_V = -2\sum_i{\Delta_i}$ in "velocity" form, where the sum 
corresponds to number of electrons.

\section{CALCULATIONS \protect\\}
\label{sec:level2}

Photoionization cross sections of O II, O III, O IV, and O V are obtained
using wavefunction expansions for residual core ions as developed in
earlier works (Nahar \cite{snn1}). Calculations are carried out 
both in non-relativistic LS coupling and in the BPRM 
approximation. 

The wavefunctions of the target or core ion were obtained from atomic
structure calculations using a scaled Thomas-Fermi model in the code 
SUPERSTRUCTURE \cite{ss}. The computation of O II photoionization cross 
sections is repeated with the same wavefunction expansion of 12 terms 
of O III of configurations $2s^22p^2$, $2s2p^3$, $2s^22p3s$, and $2p^4$ 
as in Ref. \cite{snn1}. The wavefunction expansion for O III consists 
of configurations $2s^2p$, $2s2p^2$, $2p^3$, $2s^23s$, $2s^23p$, $2s^23d$,
$2s2p3s$, and $2s2p3p$ of O IV. However, to include the fine structure
in calculating $\sigma_{PI}$ for $2s2p^3(^5S^o)$ state of O III, an 
expansion of 15 levels of O IV with configurations $2s^2p$, $2s2p^2$, 
and $2p^3$ was used. The O IV eigenfunctions consist of a 20-level 
expansion of the core ion O V with 
configurations $2s^2$, $2s2p$, $2p^2$, $2s3s$, $2s3p$, $2s3d$. The 
wavefunction expansion for O V consists of 9 states of the core ion O VI
with $1s^22s$, $1s^22p$, $1s^23s$, $1s^23p$, $1s^23d$, $1s^24s$, 
$1s^24p$, $1s^24d$, and $1s^24f$.
The calculated target level energies were replaced by the
slightly different observed 
energies to obtain somewhat more accurate positions for 
resonances that belong to Rydberg series converging on to the target
thresholds. 

The short-range bound channel correlation term in the CC expansion, 
the second term in Eq. (1), includes all possible 
(N+1)-electron configurations as in Ref. \cite{snn1} for LS coupling 
calculations. BPRM calculations are carried out for O III and O IV to
elicit fine structure details apparent in experimental data. For 
the BPRM calculations of O III, all configurations are considered with 
maximum orbital occupancies $2s^2$, $2p^4$, $3s^2$, $3p^2$, $3d^2$, while the 
K-shell remains filled. For O IV, the (N+1)-electron configurations are up
to $1s^2$, $2p^3$, $3s^2$, $3p^2$, $3d^2$, $4s^2$, $4p^2$, while K-shell is
open with a single occupancy.

Computations of $\sigma_{PI}$ are carried out using the package of
BPRM codes from the Iron Project \cite{ip2}. The present calculations
are considerably 
longer than earlier works for detailed comparison with experiments. 
They are repeated many times over, scanning with finer energy meshes as 
needed, to search for and delineate narrow resonances.  For complex ions, 
resonances are extensive and overlapping, and are easy to miss when 
energy mesh is not fine enough.  A direct comparison of theoretical and 
experimental resonant structures may not be possible if the resonances 
are narrower than the monochromatic bandwidth of the experimental beam. 
The resonances are usually convolved with a Gaussian distribution 
function with full width half maximum (FWHM) equal to the energy bandwidth 
of the monochrometer in the experiment. The average cross section, 
$\sigma_{PI}^{av}(E_i)$, at energy $E_i$ convoluted over the FWHM, $ E_o$, 
or the monochromatic bandwidth, is obtained as

\begin{equation}
\sigma_{PI}^{av}(E_i) = \int_{E_i- E_o/2}^{E_i+ E_o/2}{\sigma_{PI}(E)
{exp(-x^2)\over {\sqrt{\pi}E_o}} dE}
\end{equation}
where $x={{E-E_i}\over E_o}$. Present cross sections have been convolved
with typical energy bandwidths used in various experiments for comparison.
The photon energies are shifted slightly to match the measured 
ionization thresholds of each state exactly.

\section{RESULTS AND DISCUSSIONS\protect\\}

Photoionization cross sections of the ground and several lowest excited
states of O II, O III, O IV, and O V are obtained and compared with the 
recent precise and high-resolution measurements at the Advanced Light 
Source (ALS) at Berkeley, the University of Aarh\"{u}s, and the 
University of Paris-Sud. The resonances of detailed photoionization 
cross sections have been convolved with experimental bandwidth, as 
mentioned above, using a Gaussian distribution function to compare with 
the measured cross sections. Both the detailed and convolved 
$\sigma_{PI}$ are presented for each ion. 

Photoionization cross sections in LS coupling in general show very good 
agreement with experiment, indicating that relativistic effects are not 
significant for most states considered, especially for O II and O V. 
However, inclusion of relativistic effects is crucial for some 
metastable states for O III and O IV.

The main objective of this report is to compare the theory and experiment
for the existence and the positions of resonant structures and their
identification in the cross sections. However, the exact shape and peak 
of the resonances, as well as the background cross sections, depend 
on (i) resolution, (ii) distribution function for the beam shape, and 
(iii) fractions or mixture of states in the beam. The third point is 
difficult to determine exactly as the beam composition may vary with 
energy during measurements. With some numerical variations, the 
theoretical and experimental shapes can be brought in closer agreement, 
which however is not the main emphasis of the present work. 
The Gaussian distribution function, which peaks around the center, is 
used for convolution of theoretical cross sections. This may not 
describe exactly the distribution function of the experimental beam. 
Experimental calibration of the residual ion threshold energies is 
also an important factor for locating resonance positions.

Observed resonant features in photoionization of oxygen ions have been 
identified in Refs. \cite{als,o2b,o2a,oions}. Resonances can be 
identified using quantum defect analysis: $E_r=E_t-z^2/\nu^2$, where 
$E_r$ is the resonance energy, $E_t$ is the next core threshold energy 
in Rydbergs, $z = Z-n_c$ is the ion charge of the target where $n_c$ 
is the number of core electrons, and $\nu=n-\mu$ is the effective quantum 
number with $\mu$ being the quantum defect. $\mu$ is large for low
angular momenta, such as for an $s$-electron, and decreases with higher
angular momenta approaching near zero for $d$ orbitals in oxygen ions.

The BPRM results have many more resonances than in LS coupling due to
(i) an increment in number of Rydberg series of resonances with increased 
number of core thresholds with fine structure and (ii) relativistic 
mixing of E1 allowed and intercombination transitions. LS coupling results 
are approximately the statistical average over the fine structure.

Results for each ion are presented and discussed in separate 
subsections below.

\subsection{Photoionization of O II}

Photoionization cross sections of the ground $^4S^o$ state and the
metastable $^2P^o$, $^2D^o$ states of O II in the low energy region, 
30 - 36 eV, are shown in Fig. 1. The figure has seven panels: the lowest 
three panels, Fig. 1a,b,c, present the detailed $\sigma_{PI}$ of $^4S^o$ 
(red curve), $^2D^o$ (cyan), $^2P^o$ (green) states respectively. These 
cross sections include extensive resonances due to Rydberg series of 
autoionizing states. They become extremely narrow as the autoinizing 
states converge on to the core thresholds, $^1D$, $^1P$, $^5S^o$, etc. 
and their integrated strength is usually small. They are wiped out 
during convolution, and are not observed, as can be seen in the 
convolved cross sections of $^4S^o$, $^2D^o$, and $^2P^o$ states in 
panel d. The colors of curves identify the resonances of the states 
they belong to. Panel e presents sums the convolved cross sections 
with mixing fractions 15\% for $^2P^o$, 4\% for $^2D^o$, and 43\% for 
$^4S^o$, as deduced from experimental beam mixture in the ALS 
experiment. The summed cross sections can be compared directly with 
the $\sigma_{PI}$ measured at ALS (panel b), and at Aarh\"{u}s (panel a). 
Although the mixing proportions are somewhat different
in the two experiments the prominent features appear in both. 
The ALS cross sections have much higher resolution than those from 
Aarh\"{u}s because of the much narrower bandwidth. 
However, the ALS values are relative while the Aarh\"{u}s cross sections are 
absolute. Very good agreement can be seen between the theoretical 
and experimental features. The most significant point is that all the 
resonant features in the observed photoionization spectra have now been 
found in the calculated cross sections. Previous calculations \cite{snn1} 
used much coarser energy mesh in calculating $\sigma_{PI}$ and
missed some resonances while others were not adequately resolved.

Fig. 2 presents $\sigma_{PI}$ of the same states, the ground $^4S^o$ and 
the metastable $^2P^o$, $^2D^o$, but in the high energy region, 36 - 46 
eV. It also has seven panels: the lowest three panels (a,b,c) present
the detailed $\sigma_{PI}$ of $^4S^o$ (red curve), $^2D^o$ (cyan), 
$^2P^o$ (green) states. The extensive narrow resonance structures of
highly excited Rydberg states converging on to the various core 
thresholds are nearly wiped out during convolution, as seen in the 
convolved cross sections for $^4S^o$, $^2D^o$, and $^2P^o$ states
together in panel d. Identical colors are used for both the detailed and 
the convolved cross sections of each state. Panel c presents sum of the 
convolved $\sigma_{PI}$ with mixtures of 15\% for $^2P^o$, 4\% for $^2D^o$, 
and 43\% for $^4S^o$ as in the ALS experimental beam. The summed cross 
sections are compared with the measured $\sigma_{PI}$ at ALS (panel b) 
and Aarh\"{u}s (panel a). Very good agreement can be seen for most of 
features theoretically and experimentally. While most of the observed 
resonant features in the photoionization spectra are seen in the 
calculated cross sections, some discrepancies can be noted. The 
observed resonance at 37.5 eV, belonging to $^2D^o$ state, is shifted 
to the right (higher energy) by about 0.5 eV in the calculated spectrum. 
This may have resulted from some missing correlation configurations.
Two resonances around 37.6 and 38.9 eV are also not found in the calculated
cross sections. As these resonances were not seen in the high resolution
ALS data, it is possible that they belong to some other states in the 
mixture.

The resonance series have been identified in Refs. \cite{als,o2b,o2a}. 
For example. the resonances below 31.8 eV (threshold for $^2D^o$) in 
Fig. 1 are from $2p^3(^2P^o)$ state which can photoionize to 
$^2S$, $^2P$, and $^2D$ states with $2p$ electron going to 
$\epsilon d$ or $\epsilon s$ continua. Hence, the resonances belong 
to $2p^2(^1D)nd(^2S,^2P,^2D)$ and $2p^2(^1D)ns(^2D)$ Rydberg series of 
threshold $2p^2(^1D)$. The resonances of higher angular momentum 
in general dominate over the lower ones. The $nd$ series can be 
differentiated from $ns$ series via quantum defects and similarity in 
features. For example, of the two distinct resonance series seen in the 
low energy region of Fig. 1d, the first series (with higher peaks)  
actually represents three overlapping $nd$ series of resonances, 
$2p^2(^1D)nd(^2S,^2P,^2D)$ and the second one (lower peaks) is the
$ns$ series, $2p^2(^1D)ns(^2D)$. The table of identified resonances
are given in Refs. \cite{als,o2b}.

\subsection{Photoionization of O III}

The first measurement of photoionization cross sections for O III is 
reported by the University of Paris-Sud \cite{oions}. To compare with
the experiment, we repeat calculations to obtain high-resolution cross 
sections for the three lowest states, $2s^22p^2(^3P, ^1D, ^1S)$ of the 
ground configuration, and the metastable $2s2p^3(^5S^o)$ state of 
O III using the 23CC wavefunction expansion for the residual Boron-like 
ion O~IV of Nahar \cite{snn1}. The detailed cross sections of 
$2s^22p^2(^3P,^1D,^1S)$ states are presented in three different panels, 
a-c, of Fig. 3 ($^3P$ - red, $^1D$ - blue, and $^1S$ - green) and the
convolved cross sections are in panel d and the weighted sum in
panel e. Present detailed cross sections 
with higher resolution display much more narrow resonances 
compared to the earlier calculations. Comparison between theory and
experiment, panels e and f, shows that all the resonances observed for 
the three states are accounted for in the calculated values.

While relativistic effects are not 
important to bring out the main features of the
three $2s^22p^2(^3P,^1D,^1S)$ states, a closer examination of the
experimental results shows that there are additional features,
especially in the low energy region (see panel f). We therefore 
consider the first odd parity metastable state $2s2p^3(^5S^o)$
of O III and include fine structure through BPRM calculations. 
Prominent resonant features are found in the low energy region of
$\sigma_{PI}$ of this state that are not allowed in LS coupling, but 
are due to relativistic channel mixing. In particular the very first
resonance at $\sim$48 eV in the experimentally observed cross sections 
is due to photoionization of the high-lying metastable $2s2p^3(^5S^o)$.

We outline the differences in $\sigma_{PI}$ between the
LS coupling and the BPRM fine structure cross sections in Fig. 4. 
Figs. 4a and b present $\sigma_{PI}(^5S^o)$ in LS coupling and 
in the BPRM approximation respectively. The cross sections in the two
panels are very similar from a photon energy of 56.4 eV onwards. Whereas
$\sigma_{PI}$ is zero below this threshold in LS coupling, inclusion of 
relativistic effects in the BPRM calculations shows the region filled 
with extensive resonances with almost zero background cross section.
At 56.4 eV, $2s2p^3(^5S^o)$ of O III ionizes leaving the core ion in the 
excited $^4P$ state of O IV, which is the first threshold to the 
electron continuum. In LS coupling, $2s2p^3(^5S^o)$ state photoionizes 
only into a $^5P$ (e~+~ion) angular and spin symmetry.  
The $^5P$ symmetry can not be formed from the ground state continuum
of O IV i.e. $^2P^onl$. In other words, 56.4 eV is the ionization 
threshold for the $^5S^o$ metastable state of O III, and equals the 
sum of 47.4 eV, for ionization into the ground $^2P^o$ state, and 
8.85 eV, for excitation of the core to the $2s2p^2(^4P)$ threshold. 
No features can be formed below this threshold in LS coupling. 
However, fine structure allows ionization of the $2s2p^3(^5S^o_2)$ level 
of O III into the core ground state of O IV through relativistic mixing. 

Expanded and detailed $\sigma_{PI}$ of O III below the $^4P$ threshold of 
O IV are presented in Fig. 4c to enable comparison with experiment.
Patterns of various resonance complexes, as denoted by R$^n$ series in
the figure, can be discerned. The resonances belong to several 
Rydberg series of autoionizing levels, $^4P_{1/2,3/2,5/2}nd(^5P_{J})$ and 
$^4P_{1/2,3/2,5/2}ns(^5P_{J})$. The $^5P_{1,2,3}$ levels are allowed to
decay to $^2P^o{1/2,3/2}np(^{1,3}P_{1,2,3})$ levels via radiationless 
transitions. Decay to other levels, such as $^{1,3}D_{1,2,3}$,
$^{1,3}F_{2,3}$ are also possible through some relativistic mixing. 
However, these transitions are very weak. 

The convolved cross sections with experimental monochromatic 
bandwidth for $^5S^o$ state are shown Fig. 3d. Many theoeretically 
computed resonances are difficult to observe in the experimental data 
shown in the top panel of Fig. 3 since their width is less than
the monochromatic bandwidth of the detector in the experiment. 
A weighted sum of the ground and all metastable states included is given 
in Fig 3e. The excellent agreement between theory and experiment shows
that the theoretical O III cross sections accurately reproduce the
corresponding features in the experimental data.

\subsection{Photoionization cross sections of O IV}

Fine structure is considered in calculating photoionization cross 
sections of O III to show that large number of resonances can be
delineated. With increasing ion charge, and depending on the ionic
species, fine structure can manifest itself more prominently which is
the case for photoionization of boron-like O IV. We present $\sigma_{PI}$
for fine structure levels of the ground LS term, $2s^22p(^2P^o_{1/2,3/2})$, 
and of the first metastable term, $2s2p^2(^4P_{1/2,3/2,5/2})$ of O IV. 
Earlier works on C II, an isoelectronic partner of O IV, showed that
some resonances observed in the experiment were missing from the LS 
coupling results \cite{kjl}. Both relativistic mixing and fine structure 
splitting were found to play a role in photoionization of B-like ions
and explained the missing resonances \cite{snn2}. 

Fig. 5 presents relativistic $\sigma_{PI}$ for the levels 
$^2P^o_{1/2,3/2}$ and $^4P_{1/2,3/2,5/2}$. Cross sections for the
fine structure levels of each term show very similar features. The number 
of resonances in the BPRM results is much larger in comparison to those 
in LS coupling, as expected owing to fine structure splitting of target 
thresholds resulting in a larger number of Rydberg series of
resonances. For example, the 12 LS terms of the core ion O V correspond 
to 20 fine structure levels, which provide 8 more series of
resonances. Some addtional resonances are also produced due to 
to relativistic fine structure couplings ($\Delta$ J = 0, $\pm$ 1) not 
allowed via LS coupling.  

Figs. 5c-e show prominent resonant features, although with almost zero
background, in the low energy BPRM cross sections of $^4P$ levels. 
Extensive narrow resonances are seen to converge at about 78.7 eV,
confirming enhancement in the observed photoionization cross section
(Fig. 6c). Similar to the $^5S^o$ state of O III, channel couplings of 
fine structure levels are important in photoionization of the metastable 
$^4P$ term of O IV in forming these resonances. Without fine structure 
mixing there will be no 
cross section below 78.7 eV, the ionization threshold in LS coupling
for $^4P$ state leaving the core O V in $^3P^o$ state. At this LS 
term energy, both the BPRM and LS coupling calculations give
the same photoionzation cross sections (analagous to the $^5S^o$ state of
O III). However, fine structure levels coupling lowers the ionization 
threshold from 78.7 eV by about 10.2 eV to 68.5 eV; where 10.2 eV is the 
energy difference between the target thresholds $2s^2(^1S)$ and 
$2s2p(^3P^o)$ of O V (the O IV $^4P$ couples and photoionizes into the 
latter state). These low energy resonances can be explained as follows.
The $2s2p^2~^4P$ state can photoionize to $^4D^o,^4P^o,^4S^o$ SL$\pi$ 
symmetries of the (e~+~ion) continua. Interaction of channels with fine 
structure allows couplings with total angular momenta (SL)J$\pi$ $J\pi$ 
= $1/2^o$ and $3/2^o$. The autoionizing levels $2s2p(^3P^o)nd(^4D^o)$ 
thereby decay to $J\pi$=$1/2^o$ and $3/2^o$ continua 
$2s^2(^1S)\epsilon p(^2P^o)$ via radiation-less transitions. 

To compare these photoionization cross sections directly with the 
experimentally measured ones, presented in Fig. 6a, the cross sections 
are convolved using a Gaussian distribution function with FWHM equal to 
the monochromatic beamwidth. The $\sigma_{PI}$ of each level are then 
statistically averaged
($\sum_i{(2S_i+1)(2L_i+1)\sigma_{PI}}/\sum_i{(2S_i+1)(2L_i+1)}$) to
obtain the convolved cross sections of $^2P^o$ and $^4P$ states of O IV. 
These convolved and averaged $\sigma_{PI}$ are presented in Fig. 6c 
(red - $^2P^o$ and green $^4P$). $\sigma_{PI}(^4P)$ are 
multiplied by 9\% and $\sigma_{PI}^2P^o$ by 91\% before addition to 
obtain the weighted sum $\sigma_{PI}$. These weighted cross sections 
are given in Fig. 6b. The top panel presents the experimentally 
measured photoionization cross secitons at the University of Paris-Sud 
\cite{oions}.

Resonances in the energy region below the ground $^2P^o$ ionization 
threshold, 77.4 eV (Fig. 6), are due to the metastable $^4P$ state (green 
curve in Fig. 6c). The metastable state is relatively easy to excite as it 
corresponds to a $\Delta$ n = 0 transition ($2s^22p-2s2p^2$) and its
features are seen. The first resonance in Fig. 6 is at about 69.7 eV 
and its energy position agrees both theoretically and experimentally. 
However, some shift in positions (by $\sim$ 0.5 eV) can be seen for the 
next three resonances. The exact reason for this shift is not
clear, but a possible reason could be the effect of correlations of
configuration interactions for the $^4P$ symmetry. For the rest of the 
resonances the measured positions agree with the theoretical ones.
One exception is the measured resonance at about 79.5 eV which is not 
found either in the $^2P^o$ or the $^4P$ calculated cross sections.

The calculated peaks of resonances in Fig. 6 are seen mostly higher 
than the measured ones. These peaks may be reduced via radiative decays, 
or dielectronic recombination, if the radiative rates are comparable to 
autoionization rates. However, the radiative decay rates of the relevant 
excited core thresholds ($\approx ~ 10^8,10^9~sec^{-1}$) of O V are 
several orders of magnitude lower than typical autoionization rate 
($\approx 10^{13-14}~sec^{-1}$), and thus should cause very little 
damping, i.e., reduction in peak values. As mentioned earlier, the 
calculated peaks are affected by resolution or different energy meshes, 
Gaussian energy width, and percentage abundances of states in the beam. 
Particular mixture fractions can even eliminate resonances, as seen in 
the comparison between Fig. 6 (b) and (c). Number of resonances in 
convolved cross sections in Fig. 6c are vanished in the weighted sum in
Fig. 6b. Hence, for identification of resonances. the experimental 
spectrum should be compared with the pure convolved cross sections, 
i.e. between Fig. 6a and c, and as identified in \cite{oions}.

The most significant feature in the measured cross sections is
the prominent resonance at about 77 eV (Fig. 6a). This is a combined 
effect from the near threshold resonances of the $^2P^o$ ground state 
and converging resonances of $^4P$ levels to core threshold 
$^3P^o_{1/2}$, as seen in Fig. 5. Although narrow resonances are not 
often observed becasue of low strengths, the correspondence in energy 
positions indicates that the metastable $^4P$ photoionization is making
a contribution to the feature. The difference between the theoretical
and experimental strengths in the feature is an indicative that the
mixtures of states may not be exactly the same as quoted in the 
experiment. However, the fact that Fig. 6b exhibits the features seen 
experimentally is of potential importance in the determination of fractional
abundances of the ground and excited states in the ion source. The
background cross section can be reduced or enhanced in differenct energy
regions by taking various fractions. The fractions 91\% and 9\% used in
the present simulation may not exact match the experimental ones. 
Also, theoretical resonances are resolved with a much finer energy mesh than 
the experiment beam width, and hence appear sharper. The Gaussian 
convolution does
not broaden the features as in the experiment, which indicates that 
the experimental beam distribution function was not purely Gaussian. In
addition, the FWHM may not remain constant at all energies; a 
convolved peak can be flattened with a larger bandwidth. 

Comparison between the theoretical and experimental photoionization
cross sections in Fig. 6 highlights two of the main points in this work: 
(i) a mixture of the ground $^2P^o$ and metastable $^4P$ states in 
the experimental beam
represents the observed features; both theory and experiment
show the same resonances, and (ii) the detailed cross sections 
depends on the fractional mixture of ionic states.

\subsection{Photoionization cross sections of O V}

Photoionization cross sections of O V are obtained for the ground 
$2s^2(^1S)$ and the excited $2s2p(^3P^o,^1P^o)$ states. The detailed 
features in $\sigma_{PI}$ are shown in Fig. 7a-c: $^3P^o$ - red, 
$^1D$ - green, $^1P^o$ - blue. The resonances are more resolved than
those reported earlier \cite{snn1}.

The cross sections of the states are convolved with experimental 
bandwidth to compare with measured values at the University of 
Paris-Sud \cite{oions}. The convolved cross sections are presented in 
Fig. 7d, keeping the original colors for each state for easy 
identification. For comparison with experiment, thes convolved cross 
sections are summed with weighting factors 75\% for $^1S$, 23\% for 
$^3P^o$, and 2\% for $^1P^o$, whereas the fractions in the 
experimental beam as reported in Ref. \cite{oions} are 65\% for the
ground state, $^1S$, and 35\% for excited $^3P^o$ state. The weighted 
sum in Fig. 7e is compared with the measured spectra in Fig. 7f. 
The reason for inclusion of $2s2p(^1P^o)$ cross sections is similarity
of features. Although the measured features in the cross sections below 
103.6 eV, the ionization threshold for $2sp2(^3P^o)$ state, appear as 
small kinks in Fig. 7f, they correspond to the resonant features 
appearing at the same energy positions of $2s2p(^1P^o)$ state of O V.

A clear rise seen in the background cross sections at about 103.6 eV 
corresponds to the $^3P^o$ ionization threshold, and at about 113.85 
eV to $^1S$ ionizaton threshold, as noted in 7e and 7f. However, 
differences in the background cross sections are found. Since present
mixture includes higher percentage of the $^1S$ ground state, similar
enhancement is expected in the weighted average. The calculated resonant
peaks are also higher than the observed ones. Discussion regarding the
differences in peaks is made above. However, the main objective is
met, that is, all observed resonant features, shapes and energy positions 
are reproduced in the calculated spectra. Identifications of the
resonances are given in Ref. \cite{oions}.

\section{CONCLUSION}

Photoionization cross section of oxygen in several stages of ionization 
have been obtained for the ground and low lying excited states. These
cross sections are compared with measured cross sections at three
different experimental set-ups.

While the experimental measurement includes combined features 
of resonances belonging to different Rydberg series of states, and of 
a mixture of different states, theoretical work can evaluate them for 
individaul states. 

Certain features in the measured cross sections are not seen prominently 
or are missing in previous calculations. Present report shows that the 
missing features are due to (i) lower resolution of previous calculations 
or (ii) due to relativistic fine structure effects, not included in 
the previous calculations. For most cases of this relatively low Z
element, $LS$ coupling calculations are often adequate. However, for
some metastable states, fine structure is crucial for the low energy
resonances which allowed only the couplings of fine structure levels.
It is the fine structure components, rather than the relativistic effects,
that is important for the formation of these high peak resonances with
nearly zero backgound. Inclusion of relativistic effects will certainly
be more exact and accurate way to carry out the calculations.

\acknowledgments

I would like to thank Anil Pradhan for discussions and suggestions.
This work was partially supported by the National Science Foundation
and the NASA Astrophysical Theory Program. The computational work was
carried out at the Ohio Supercomputer Center.

%***
%***  E n d   o f   p a g e   1   o f   g a l l e y - m o d e   o u t p u t
%***

\begin{figure}
\centering
\psfig{figure=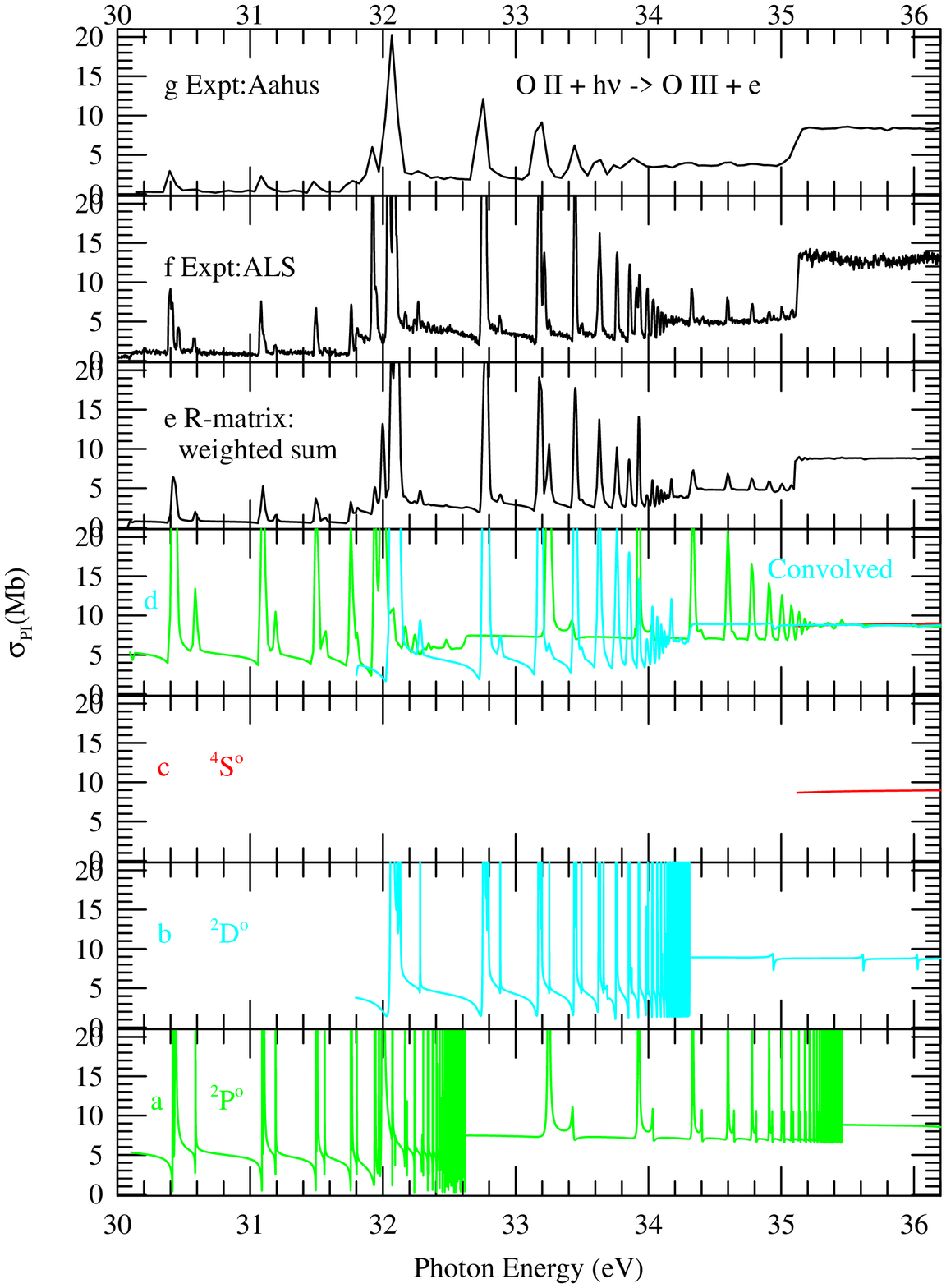,height=15.0cm,width=18.0cm}
\caption{Photoionization cross sections $\sigma_{PI}$ of ground $^4S^o$ 
and metastable $^2P^o$, $^2D^o$ states of O II in low energy region,
30 - 36 eV: Panels a, b, c - detailed $\sigma_{PI}$ of  $^2P^o$ (green), 
$^2D^o$ (cyan), and $^4S^o$ (red) states respectively; d - the 
same cross sections, but are convolved with experimental monochrormeter 
bandwidth: e - sum of the convolved cross sections weighted by 
experimental beam mixtures to compare with measured $\sigma_{PI}$; 
f, g - measured  $\sigma_{PI}$ by Covington et al. (2001) and by Kjeldsen 
et al. (2002), respectively.}
\end{figure}

\begin{figure}
\centering
\psfig{figure=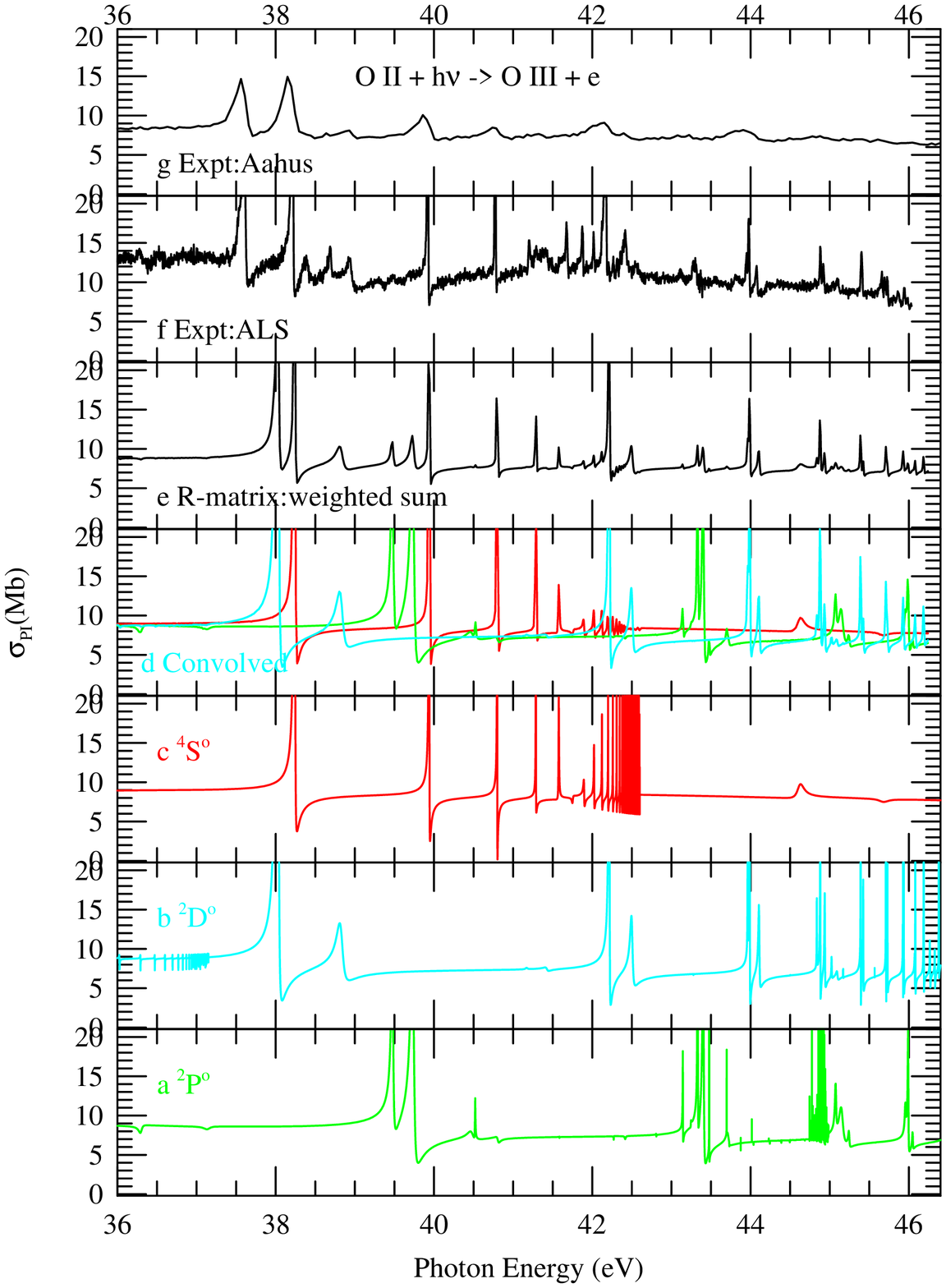,height=15.0cm,width=18.0cm}
\caption{Photoionization cross sections $\sigma_{PI}$ of ground $^4S^o$
and metastable $^2P^o$, $^2D^o$ states of O II in the higher energy region.  
36 - 46 eV: Panels a, b, c - detailed $\sigma_{PI}$ of  $^2P^o$ (green),
$^2D^o$ (cyan), and $^4S^o$ (red) states respectively; d - the
same cross sections, but are convolved with experimental monochrormeter 
bandwidth: e - sum of the convolved cross sections weighted by 
experimental beam mixtures to compare with measured $\sigma_{PI}$; 
f, g - measured  $\sigma_{PI}$ by Aguilar et al. (2003) and by Kjeldsen 
et al. (2002), respectively.}
\end{figure}

\begin{figure}
\centering
\psfig{figure=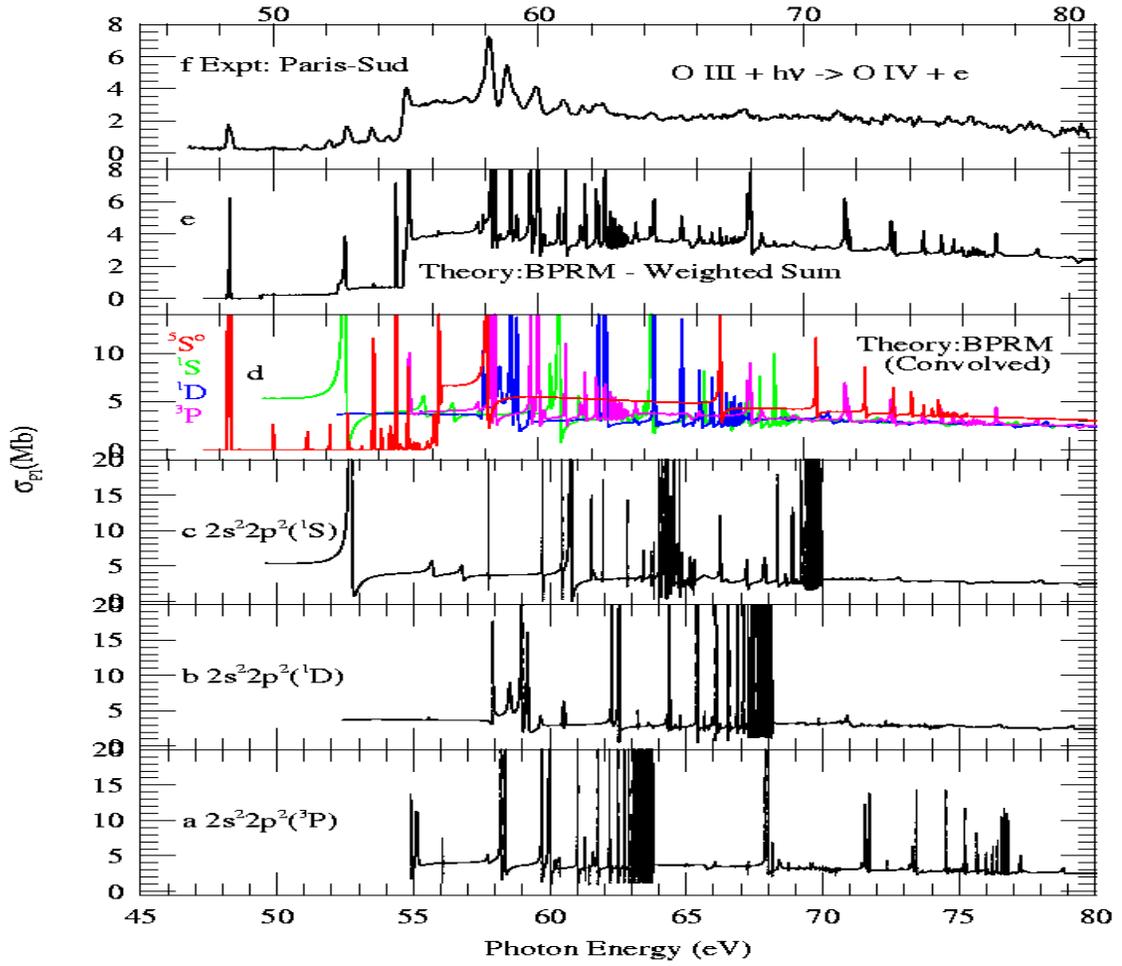,height=15.0cm,width=18.0cm}
\caption{Photoionization cross sections $\sigma_{PI}$ of O III: 
Panels a, b, c - $2s^22p^2(^3P,^1D,^1S)$ states; d - convolved 
cross sections of $^3P,^1D,^1S,^5S^o$; e - weighted sum of convolved 
cross sections; f - experimental cross sections of the combined states
at University of Paris-Sud by Champeaux et al. (2003).}
\end{figure}

\begin{figure}
\centering
\psfig{figure=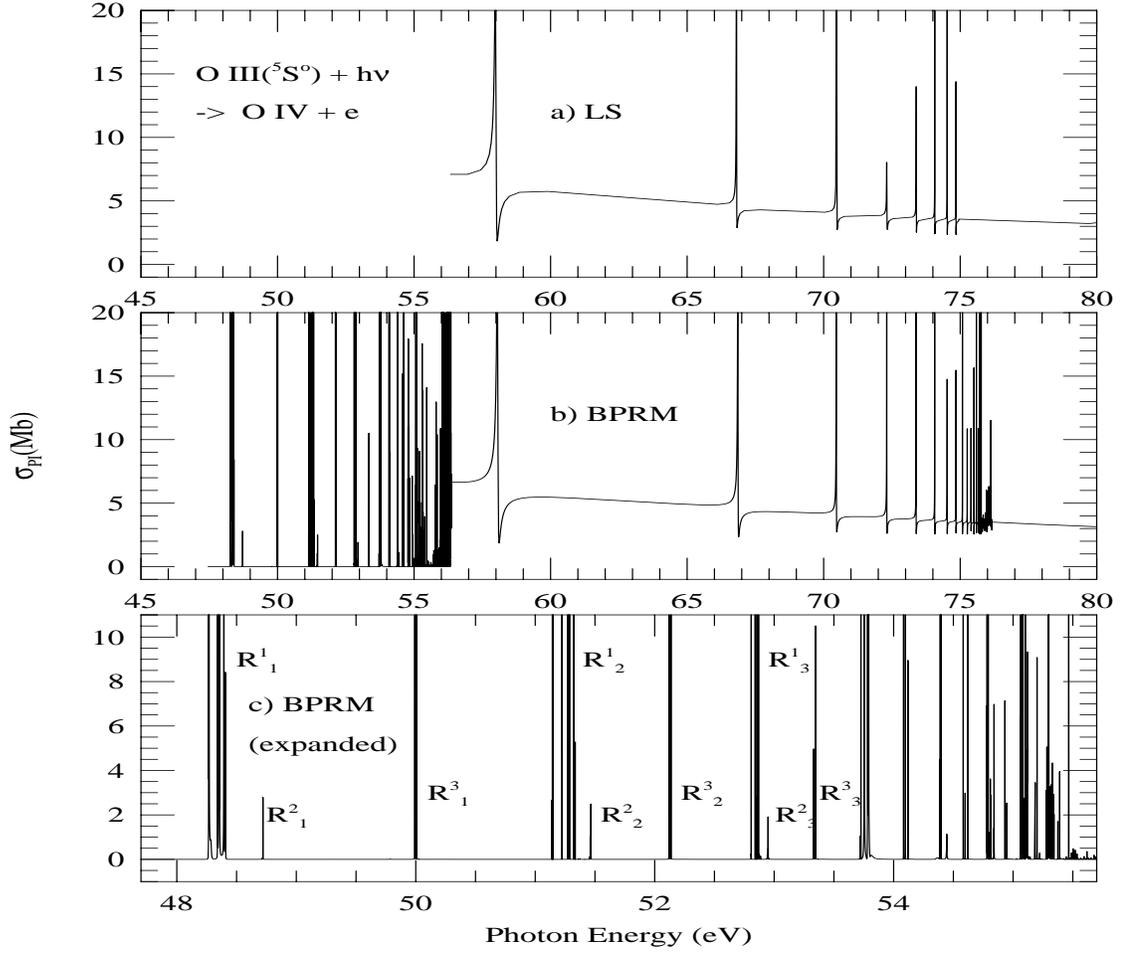,height=15.0cm,width=18.0cm}
\caption{Photoionization cross sections $\sigma_{PI}$ of excited 
metastable $2s2p^3(^5S^o_2)$ state of O III: Panel a - in LS coupling; b - 
in BPRM approximation; c - $2s2p^2~^4Pnl(R)$ series identificatioin in low 
energy region.}
\end{figure}

\begin{figure}
\centering
\psfig{figure=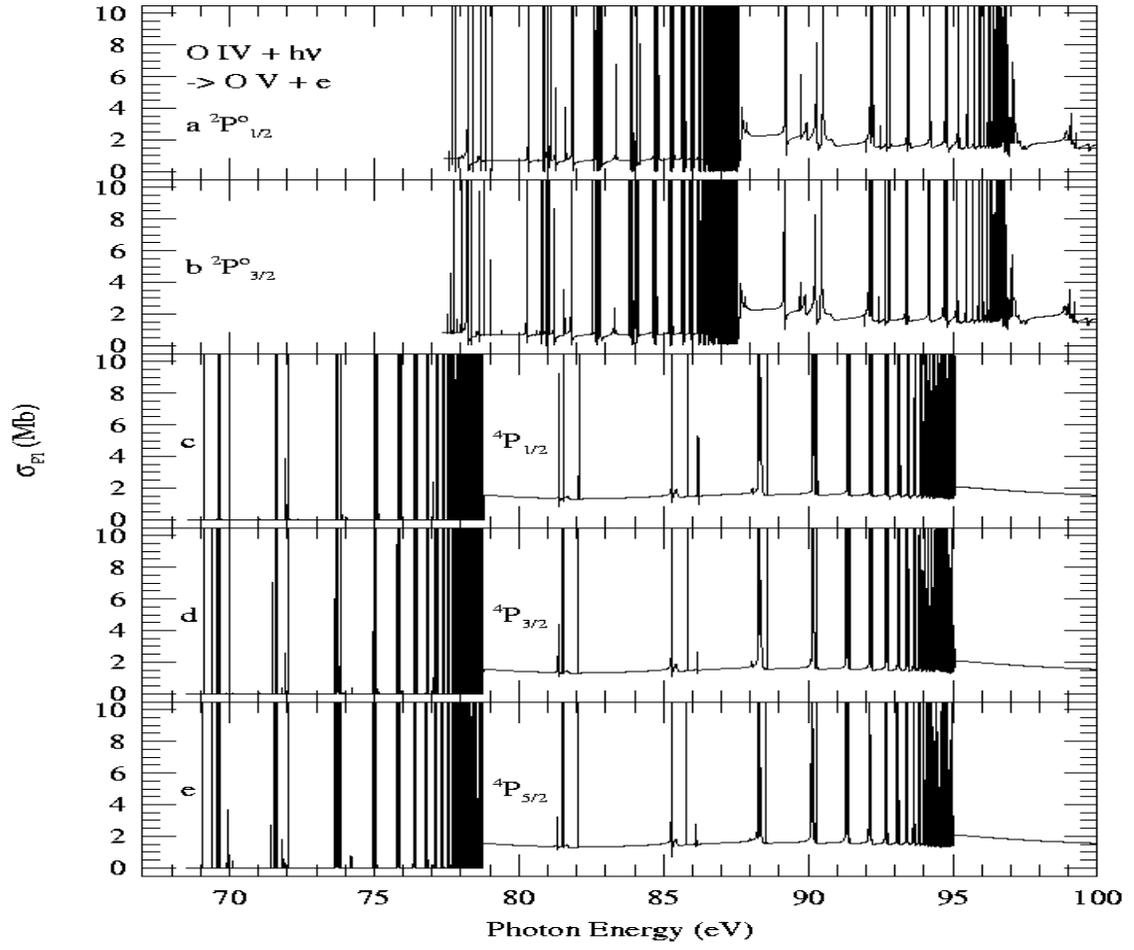,height=15.0cm,width=18.0cm}
\caption{Photoionization cross sections $\sigma_{PI}$ of O IV in BPRM
approximation: a, b - $^2P^o_{1/2,3/2}$ levels of ground 
configuration, $2s^22p$; c, d , f - excited $2s2p^2(^4P_{1/2,3/2,5/2})$
levels.} 
\end{figure}

\begin{figure}
\centering
\psfig{figure=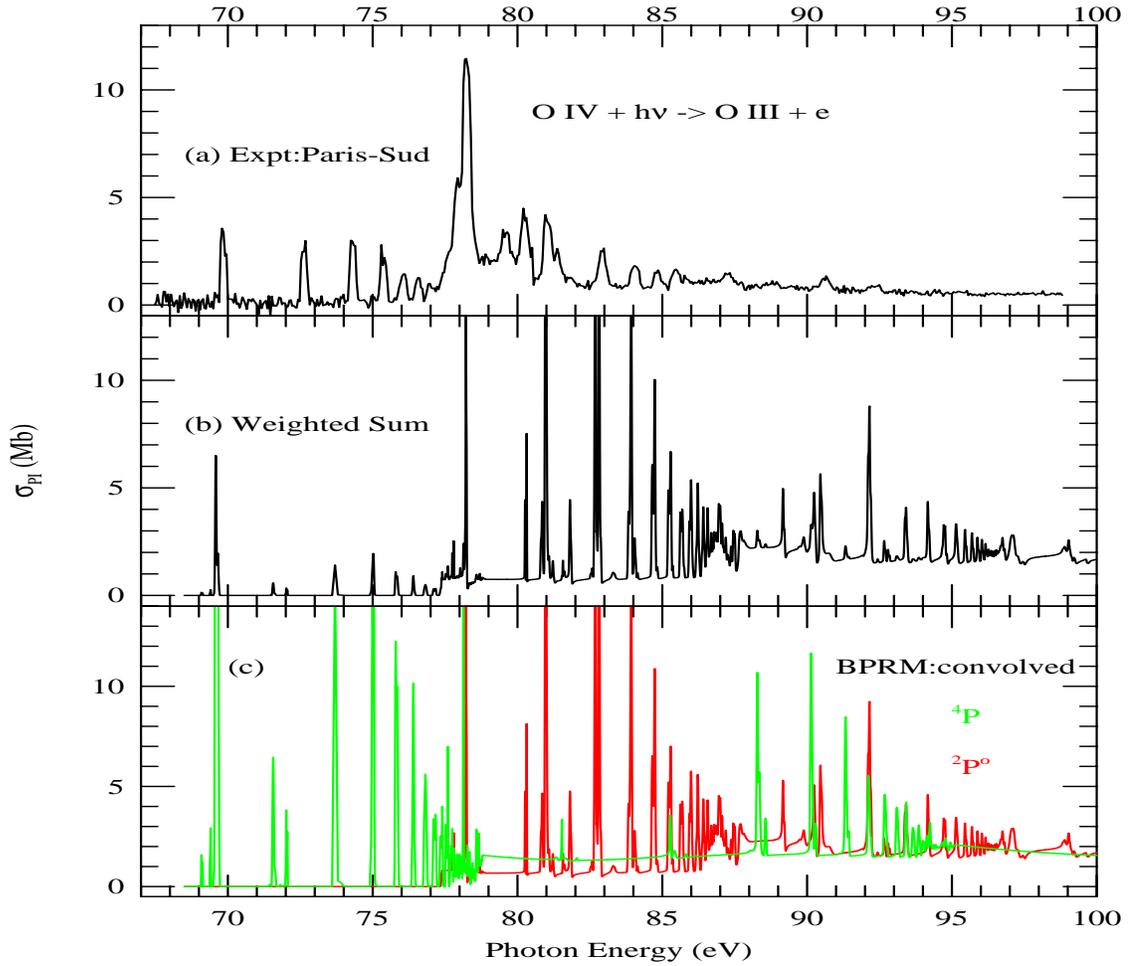,height=15.0cm,width=18.0cm}
\caption{Comparison of photoionization cross sections $\sigma_{PI}$ of O IV 
with experiment: c - red and green curves are statistical sum of convolved 
cross sections of levels $^2P^o_{1/2,3/2}$ and $^4P_{1/2,3/2,5/2}$ 
respectively; b - weighted sum of the cross sections; c - measured cross 
sections at University of Paris-Sud by Champeaux et al. (2003).}
\end{figure}

\begin{figure}
\centering
\psfig{figure=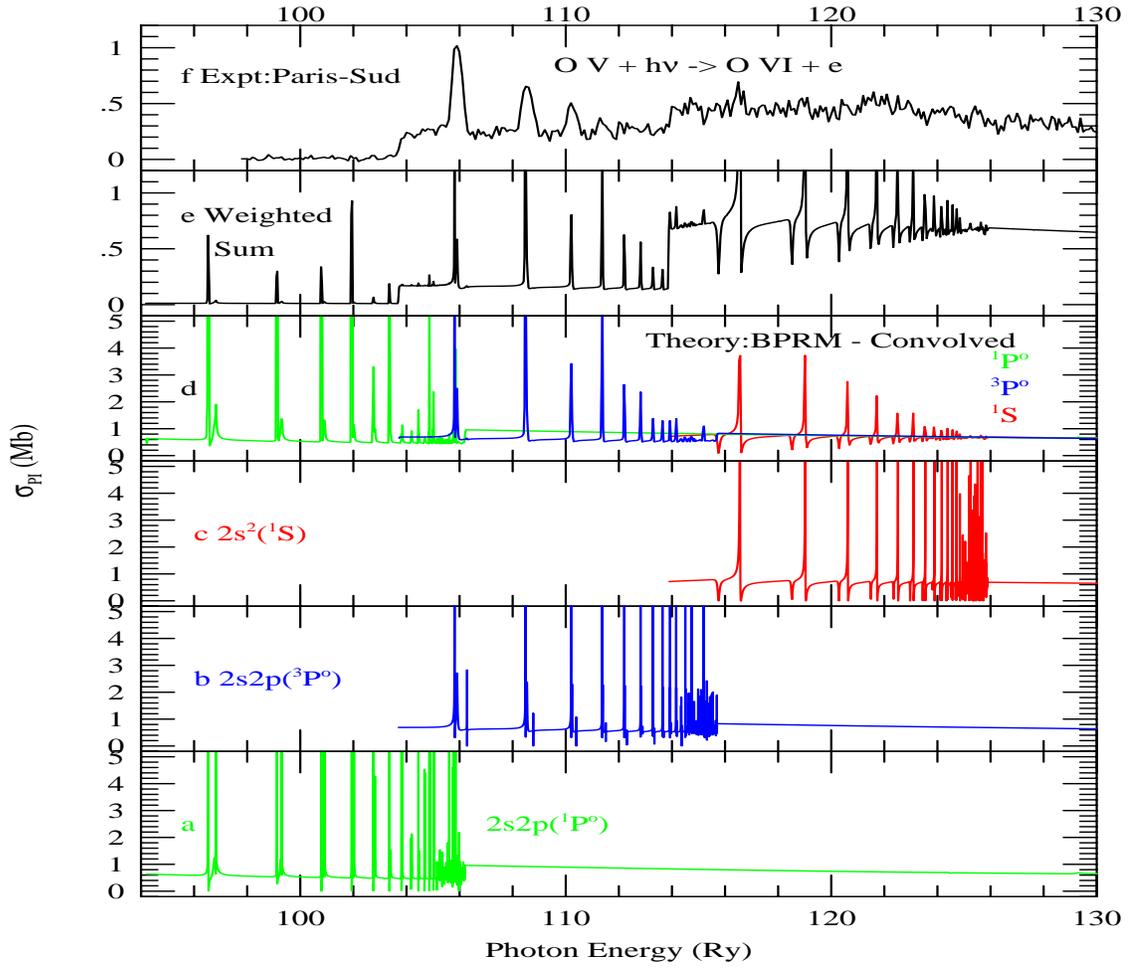,height=15.0cm,width=18.0cm}
\caption{Photoionization cross sections $\sigma_{PI}$ of O V: a, b, c - 
Detailed $\sigma_{PI}$ of ground $2s^2(^1S)$ and excited $2s2p(^3P^o,^1P^o)$ 
states; d - convolved cross sections, e - weighted sum of convolved cross 
sections; f - measured cross sections at University of Paris-Sud by
Champeaux et al. (2003).}
\end{figure}

%***
%***  E n d   o f   p a g e   6   o f   g a l l e y - m o d e   o u t p u t
%***

\end{document}